\documentclass[12pt]{iopart}
\usepackage{graphicx}
\usepackage{iopams}
\usepackage[colorlinks=true,linktocpage=false,citecolor=blue]{hyperref}

\def\eps{\epsilon}
\def\wh{\widehat}

\begin{document}
\title{A Semi-relativistic Model for Tidal Interactions in BH-NS Coalescing
  Binaries}

\author{V Ferrari$^1$, L Gualtieri$^1$ and F Pannarale$^1$}
\address{$^1$ Dipartimento di Fisica ``G.Marconi'', Sapienza
Universit\` a di Roma \\ and Sezione INFN ROMA1, piazzale Aldo Moro 2,
I-00185 Roma, Italy}

\begin{abstract}
We study the tidal effects of a Kerr black hole on a neutron star in
black hole-neutron star binary systems using a semi-analytical
approach which describes the neutron star as a deformable ellipsoid.
Relativistic effects on the neutron star self-gravity are taken into
account by employing a scalar potential resulting from relativistic
stellar structure equations.  We calculate quasi-equilibrium sequences
of black hole-neutron star binaries, and the critical orbital
separation at which the star is disrupted by the black hole tidal
field: the latter quantity is of particular interest because when it
is greater than the radius of the innermost stable circular orbit, a
short gamma-ray burst scenario may develop.
\end{abstract}

\section{Introduction}
During the past decades, theorists have been modelling various kinds
of double compact objects since (1) they are among the most promising
gravitational wave sources to be detected by ground-based and
space-based laser interferometers \cite{Detectors} and (2) they have
also been invoked as possible engines of short gamma-ray bursts
\cite{GRB1} (see also \cite{GRB,LeeGRB}) in the case of black
hole-neutron star (BH-NS) and neutron star-neutron star (NS-NS)
mergers. The remnants of both kinds of mergers, in fact, may result in
a black hole with negligible baryon contamination along its polar
symmetry axis and surrounded by a hot massive accretion disk: before
the disk gas is accreted to the black hole, intense neutrino fluxes
are emitted which, through energy transfer, trigger a high-entropy gas
outflow off the surface of the accretion disk (``neutrino wind''); at
the same time, energy deposition by $\nu\bar\nu$ annihilation in the
baryon-free funnel around the rotation axis, powers relativistically
expanding $e^\pm\gamma$ jets which can give rise to gamma-ray bursts
\cite{GRB1}. The fate of BH-NS binaries, in particular, depends on the
relative values of $r_{ISCO}$, the radius of the innermost stable
circular orbit, and $r_{tide}$, the orbital separation at which the
tidal disruption of the star by the BH occurs: if $r_{ISCO}<r_{tide}$
the star is disrupted and then swallowed by the BH, otherwise it is
swallowed without disruption. This is therefore a crucial issue for
the SGRB mechanism we have described: only if $r_{ISCO} < r_{tide}$
the merger may result in the black hole + hot massive accretion disk +
baryon-free funnel SGRB scenario.

A significant effort in studying BH-NS coalescing systems has been
undertaken in order to model them as gravitational wave sources and to
understand gamma-ray burst engines. Unfortunately, BH-NS as well as
BH-BH binaries (as opposed to NS-NS binaries) have not been observed
yet and therefore it is not currently possible to infer their
properties from observational data. Thus, in modelling the behaviour
of such binaries and in making predictions about them, one has to rely
on binary evolution and population synthesis models. Several
approaches addressing different issues have been employed to study
coalescing binaries. Post-Newtonian (PN) studies of compact binaries
(\cite{Blanchet} and references therein), for example, typically
approximate the constituents of the binary as point sources. PN
expansions may not converge rapidly enough in the strong-field region
and thus are indicated for the inspiral phase: this has lead to a
strong effort in stitching together PN studies with other methods that
are more indicated for addressing the merging and ringdown, e.g.
\cite{BertiIyerWill}. Other approaches deal with finite size effects
due to at least one of the binary constituents; many of them assume
Newtonian gravity in some or all aspects of the calculation. Finite
size-effects are connected to tidal interactions between the binary
constituents: these interactions are present well before the final
phases of the coalescence and have been studied in the literature
using various approaches and addressing several issues. We refer the
reader to \cite{Christian-articolo} for an overview on the literature.

Nowadays, the mainstream strategy in studying BH-NS binaries is to
develop fully relativistic codes, as is being done over the years by
several groups (see \cite{GR_eg},\cite{TBFS08} for recent
work). However the high computational cost of simulations of compact
binaries makes it desirable to have frameworks to study certain
features and dynamical regimes which admit the use of approximations,
thus allowing a drastic reduction of the computational resources
required. Motivated by this reason and by the intent of investigating
SGRB engines, in this paper we focus on BH-NS binaries and develop a
semi-relativistic model in order to describe the behaviour of a NS
undergoing tidal interactions with its BH companion.

Our starting point is the model developed in
\cite{WL2000} by Wiggins and Lai:
\begin{itemize}
\item it is essentially the \emph{affine model} approach of Carter,
  Luminet and Marck \cite{CL85},\cite{LM85}, i.e. the NS is treated as
  a \emph{Newtonian} extended object which responds to its
  self-gravity, to its internal pressure forces and to the
  \emph{relativistic} BH tidal field under the constraint that its
  shape is always that of an ellipsoid (hence the name ``affine
  model''); this constraint descends from the fact that ellipsoidal
  figures are solutions to the dynamical equations describing a
  Newtonian incompressible star responding to the linear terms of a BH
  tidal field (see \cite{EFE} for an extensive explanation).
\item NSs on equatorial circular orbits around a BH are considered and
  the quasi-equilibrium sequences of a \emph{Newtonian polytropic}
  NS\footnote{Wiggins and Lai also consider white dwarfs, but we are
  not interested in white dwarfs in this paper.} interacting with a
  Kerr BH are determined from large separation until the star tidal
  disruption: the critical orbital radius at which the tidal
  disruption occurs is compared to the radius of the ISCO in the
  context of SGRB engines.
\end{itemize}
A similar approach has been used in \cite{LS76}.

In this paper we improve their model in two directions by
\begin{enumerate}
\item including general relativistic effects in the response of the NS
  to the BH tidal field
\item using a general, barotropic equation of state (EOS) for the NS.
\end{enumerate}
Obviously, both points arise from the desire of building a model which
has as many ``realistic'' features as possible; however there is
more. The NS disruption, which may lead to the SGRB mechanism
previously described, occurs when the BH tidal force starts prevailing
on the NS self-gravity; this condition may be approximately expressed
as $\alpha M_{NS}/R^2_{NS} \simeq M_{BH}R_{NS}/r_{tide}^3$, where
$\alpha$ is a dimensionless coefficient, $M_{NS}$ and $R_{NS}$ are the
NS mass and radius, and $M_{BH}$ is the BH mass.  $r_{tide}$ hence
depends essentially on two parameters: the mass ratio
$q=M_{BH}/M_{NS}$ and the NS compactness at equilibrium
$C=M_{NS}/R_{NS}$. The main advantage of Wiggins and Lai's model as
opposed to many fully relativistic formulations, is that it allows one
to choose large values of $q$.  On the other hand, the weak point in
ellipsoidal models in general, is that Newtonian self-gravity is
adopted, and this choice is inappropriate to describe very compact
stars.

In our approach we improve this point by approximating the effects of
general relativistic gravity by means of an effective scalar
gravitational potential $\Phi_{TOV}$. This potential (see
\ref{app:pseudoTOV}) stems from the Tolman-Oppenheimer-Volkoff (TOV)
stellar structure equations in General Relativity, and has been
adopted and tested in the context of stellar core collapse and
post-bounce evolution simulations:
\begin{itemize}
\item in \cite{RamppJanka2002} Rampp and Janka present the
  \textsc{Vertex} code for supernova simulations; in order to
  approximate relativistic gravity, this code makes use of the
  generalised potential in place of the usual Newtonian potential in
  all Newtonian hydrodynamics equations
\item in \cite{Liebendorfer05} Liebend{\"o}rfer et al. perform a
  comparison between results of the \textsc{Vertex} and
  \textsc{Agile-BoltzTran} codes, the latter being a fully
  relativistic (1D) hydrodynamics code; it is shown that both codes
  produce qualitatively very similar results except for some small
  (but growing) quantitative differences occurring in the late
  post-bounce evolution
\item in order to achieve a better agreement than that reported in
  \cite{Liebendorfer05}, different improvements of the aforementioned
  effective relativistic potential are explored and tested by Marek
  and collaborators in \cite{Dimmelmeier2006}; the Newtonian equations
  of hydrodynamics remain untouched.
\end{itemize}
The way we apply the main idea of
\cite{RamppJanka2002}-\cite{Dimmelmeier2006} in the present,
completely different, context is the following: the Eulerian
hydrodynamics equations --- which govern the star in the affine model
and which are Newtonian --- are left formally unchanged, but the
pressure profile of the star at equilibrium appearing in them is built
with the TOV equations and gravitational potential used is
$\Phi_{TOV}$.

Furthermore, we extend the affine model to include general barotropic
equations of state (see $\S$ \ref{sec:model}).

In this paper we concentrate on calculating $r_{tide}$ with our
approach.  As a test of the model we reproduce the fully relativistic
results obtained in \cite{TBFS08}, where the tidal disruption limit of
binaries containing a Schwarzschild BH and a polytropic relativistic
NS is calculated. We subsequently compute $r_{tide}$ for binaries
composed of a Kerr black hole and a NS, modeled with more realistic
equations of state.

The paper is organized as follows. In Section \ref{sec:model} we
describe the basic equations of the affine model suitably modified as
explained above; in Section \ref{sec:results} we test our model and
present the results of the numerical simulations; in Section
\ref{sec:conclusions} we draw the conclusions.

\section{The Model}\label{sec:model}
In this section we build our BH-NS model whose essential features are:
\begin{itemize}
\item the NS moves in the Kerr BH tidal field along the BH
  timelike geodesics
\item it maintains an ellipsoidal shape; more precisely it is a
  Riemann-S type ellipsoid, i.e. its spin and vorticity are parallel
  and their ratio is constant (see \cite{EFE})
\item its equilibrium structure is determined using the equations of
  General Relativity, its dynamical behaviour is governed by Newtonian
  hydrodynamics improved by the use of an effective relativistic
  potential
\item its EOS can be any (tabulated or analytic) barotropic EOS.
\end{itemize}
In the spirit of \cite{WL2000}, the equations for the NS will be
written in the principal frame, i.e. the frame associated to the
principal axes of the stellar ellipsoid, which is set up so that ``1''
denotes the direction along the axis that tends towards the BH, ``2''
is the direction along the other axis that lies in the orbital plane
and ``3'' is associated with the direction of the axis orthogonal to
the orbital plane.

\subsection{Overview of the Assumptions}
\label{ssec:assumptions}
We shall assume that the star moves on an equatorial circular geodesic
around the BH and neglect tidal effects on the orbital motion. We are
thus working in the \emph{tidal approximation}, i.e. we assume that:
\begin{itemize}
\item $M_{BH}\gg M_{NS}$ (where $M_{BH}$ is the black hole mass and
  $M_{NS}$ the star mass)
\item the stellar deformations do not influence its orbital motion.
\end{itemize}
We neglect the perturbation that the star induces on the BH.

According to the affine model, the surfaces of constant density inside
the star form self-similar ellipsoids and the velocity of a fluid
element is a linear function of the $x_i$'s (the coordinates in the
principal frame) \cite{LRS}. This allows one to reduce the infinite
degrees of freedom of the stellar fluid to five dynamic variables and
to deal with a (finite) set of ordinary differential equations
governing the dynamics of the star. The five fluid variables are the
three principal axes of the ellipsoid ($a_i$, $i=1,2,3$) and two
angles $\phi$ and $\lambda$ which are defined by the differential
relations
\begin{eqnarray}
\frac{d\phi}{d\tau} = \Omega,\qquad
\frac{d\lambda}{d\tau} = \Lambda,
\label{firsord}
\end{eqnarray}
where $\tau$ is the NS proper time, $\Omega$ is the ellipsoid angular
velocity measured in the parallel-transported frame associated with
the star centre of mass and $\Lambda$ characterises the internal fluid
motion by
\begin{eqnarray}
\Lambda = −\frac{a_1a_2\zeta}{a^2_1+a^2_2}
\end{eqnarray}
where $\zeta$ is the (uniform) vorticity along the $z$-axis in the
frame corotating with the ellipsoid.  Given these assumptions, the
Lagrangian governing the internal (``I'') dynamics of the star may be
written as
\begin{eqnarray}
\label{Linner}
\mathcal{L}_I = \mathcal{L}_T+\mathcal{L}_B,
\end{eqnarray}
where ``T'' stands for ``tidal'' and ``B'' for ``body''. In the next
two sections we will write down both terms explicitly.

\subsection{BH-NS Tidal Interaction}
$\mathcal{L}_T$ may be written as
\begin{eqnarray}
\label{L_T}
\mathcal{L}_T =  -\int d^3x\rho(\mathbf{x})\Phi_{tide}(\mathbf{x}) =
-\frac{1}{2}c_{ij}I_{ij},\qquad i,j=1,\ldots, 3,
\end{eqnarray}
where $c_{ij}$ and $I_{ij}$ are the components of the BH tidal tensor
and of the inertia tensor of the star. We recall that for a point mass
orbiting a Kerr BH on an equatorial circular orbit (the NS centre of
mass in our case) the energy and the $z$-orbital angular momentum per
unit mass are
\begin{eqnarray}
E = \frac{r^2-2M_{BH}r+a\sqrt{M_{BH}r}}{r\sqrt{P}},\qquad
\label{def:Lz}
L_z = \frac{\sqrt{M_{BH}r}(r^2-2a\sqrt{M_{BH}r}+a^2)}{r\sqrt{P}}
\end{eqnarray}
where $a$ is the black hole spin aligned with the orbital angular momentum,
\begin{eqnarray}
P = r^2-3M_{BH}r+2a\sqrt{M_{BH}r}
\end{eqnarray}
and out of $E$ and $L_z$ one builds the constant
\begin{eqnarray}
K = (aE-L_z)^2.
\end{eqnarray}
The components of  the tidal tensor for a Kerr spacetime are then
\begin{eqnarray}
\nonumber
&c_{11} = \frac{M_{BH}}{r^3}\left[1-3\frac{r^2+K}{r^2}\cos^2(\Psi-\phi)
\right],
&c_{22} = \frac{M_{BH}}{r^3}\left[1-3\frac{r^2+K}{r^2}\sin^2(\Psi-\phi)
\right]\\ 
&c_{12} = c_{21} = \frac{M_{BH}}{r^3}\left[ -\frac{3}{2}\frac{r^2+K}{r^2}
  \sin 2(\Psi-\phi)\right],
&c_{33} = \frac{M_{BH}}{r^3}\left( 1+3\frac{K}{r^2}\right)
\label{cij:1}
\end{eqnarray}
where $r$ is the orbital separation, $\Psi$ is an angle governed by
\begin{eqnarray}
\frac{d\Psi}{d\tau} = \frac{E(L_z-aE)+a}{r^2+K},
\end{eqnarray}
which identifies the parallel-transporting frame associated with the
star centre of mass \cite{LM85} and $\phi$ is the angle given in
(\ref{firsord}), that brings this frame into the principal frame by a
rotation around an axis orthogonal to the orbital plane and passing
through the star centre.

The components of the tensor of inertia $\mathbf{I}$ appearing in (\ref{L_T})
are defined as
\begin{eqnarray}
\label{def:I}
I_{ij} = \int d^3x\rho x_ix_j,
\end{eqnarray}
where $\rho$ is the mass density distribution. In the principal frame,
the tensor $\mathbf{I}$ is diagonal and takes the form
\begin{eqnarray}
\label{Iij}
{\mathbf I}=\wh{\mathcal{M}}\textrm{diag}\left(\frac{a_i}{R_{NS}}\right)^2,
\end{eqnarray}
where $R_{NS}$ is the isolated NS radius at (spherical) equilibrium,
the $a_i$'s indicate the lengths of the principal axes of the stellar
ellipsoid and $\wh{\mathcal{M}}$ is the star scalar quadrupole moment
at spherical equilibrium (in isolation), i.e.
\begin{eqnarray}
\label{def:momquad}
\wh{\mathcal{M}} = \frac{1}{3}\int\hat{x}^i\hat{x}^i\hat{\rho}d^3\hat{x}=
   \frac{4\pi}{3}\int_0^{R_{NS}}\hat{r}^{4}\hat{\rho}d\hat{r}~.
\end{eqnarray}
Hereafter, the carets ($\,\hat{\;}\,$) denote variables referring to
the isolated star at equilibrium. Notice that at equilibrium, as
expected, the tensor of inertia and the scalar quadrupole moment are
related by $\textrm{Tr}(\mathbf{I})\equiv 3\mathcal{\wh{M}}$.

\subsection{Fluid Terms}
Following \cite{WL2000} we write
\begin{eqnarray}
\label{Lbody}
\mathcal{L}_B = T_I - U - V,
\end{eqnarray}
where $T_I$ is the kinetic energy of the star internal
(i.e. non-orbital) motion, $U$ is the internal energy of the stellar
fluid and $V$ is the star self-gravity potential.

In general the internal kinetic energy of a body is given by
\begin{eqnarray}
\label{T_I}
T_I = \int d^3x\frac{1}{2}\rho u^2,
\end{eqnarray}
where $\vec{u}$ is the velocity field of its internal motions which,
for a Riemann-S type ellipsoid, is \cite{EFE} $\vec{u} =\vec{u}_s +
\vec{u}_e,$ where
\begin{eqnarray}
\vec{u}_s = \left(\frac{a_1}{a_2}\Lambda - \Omega\right)x_2\vec{e}_1
+\left(-\frac{a_2}{a_1}\Lambda + \Omega\right)x_1\vec{e}_2
\end{eqnarray}
is the spin velocity (the speed of the fluid due to rotation), and
\begin{eqnarray}
\vec{u}_e = \frac{\dot{a}_1}{a_1}x_1\vec{e}_1
+\frac{\dot{a}_2}{a_2}x_2\vec{e}_2+\frac{\dot{a}_3}{a_3}x_3\vec{e}_3
\end{eqnarray}
is the expansion/contraction velocity of the ellipsoid.  Substituting
these expressions in (\ref{T_I}) we find
\begin{eqnarray}
\label{T_I=sum}
T_I &=& 
\sum_i \frac{1}{2}\left(\frac{\dot{a}_i}{a_i}\right)^2 \int d^3x \rho x_i^2
\nonumber
\\
&+& \frac{1}{2}\left[\left(\frac{a_1}{a_2}\Lambda-\Omega\right)^2
\int d^3x\rho x_2^2 + \left(\Omega-\frac{a_2}{a_1}\Lambda\right)^2\int d^3x\rho
x_1^2\right].
\end{eqnarray}
Since in this paper we are dealing with general barotropic EOS, the
integrals appearing in $T_I$ will have to be calculated numerically.

The internal energy of the stellar fluid $U$ is defined as the volume integral
of the fluid energy density:
\begin{eqnarray}
\label{def:U}
U = \int\epsilon d^3x.
\end{eqnarray}
For any barotropic EOS, $U$ is related to the pressure
integral\footnote{ Note that in the affine approximation
$\rho={\hat\rho}/{a_1a_2a_3}$, where $\hat\rho$ is the mass density
for the star spherical equilibrium configuration. In this approach
$dM$ is unchanged while $d^3x=(a_1a_2a_3)/R_{NS}^3d^3\hat{x}$.}
\begin{eqnarray}
\label{def:Pi}
\Pi = \int P(\rho) d^3x = 4\pi \frac{a_1a_2a_3}{R_{NS}^3}\int_0^{R_{NS}}P\left(
  \frac{\hat{\rho}}{a_1a_2a_3}\right)\hat{r}^2d\hat{r}
\end{eqnarray}
by the following differential relation \cite{CL85}
\[
dU= \sum_i \frac{\Pi}{a_i} d a_i;
\]
this equation is needed in order to write down the Lagrange
equations. Finally, the self-gravity potential in general terms is
defined as \cite{EFE}
\begin{eqnarray}
\label{def:V}
V = \frac{1}{2}\hat{V}R_{NS}\sum_ia_i^2\int_0^\infty
\frac{du}{(a_i^2+u)\Delta(u)},
\end{eqnarray}
where
\[
\Delta(u) = \sqrt{(a_1^2+u)(a_2^2+u)(a_3^2+u)},
\]
$\hat{V}$ is the self-gravitational energy for the star at spherical
equilibrium
\begin{eqnarray}
\label{def:V1}
\hat{V} = \textrm{Tr}\hat{V}_{ij}
\qquad\hbox{where}\qquad
\hat{V}_{ij} = -\int dM \partial_i(\Phi)\hat{x}_j,
\end{eqnarray}
and $\Phi$ is the gravitational potential. In spherical coordinates,
the trace of $\hat{V}_{ij}$ is
\begin{eqnarray}
\label{def:hatV}
\hat{V}=-4\pi \int_0^{R_{NS}} \frac{d\Phi}{d\hat{r}}\hat{r}^3\hat{\rho} d\hat{r}.
\end{eqnarray}
This expression must be evaluated numerically for the chosen
barotropic EOS and gravitational potential $\Phi$. In the case of a
polytropic equation of state and for a Newtonian gravitational
potential, the integral (\ref{def:hatV}) can be performed analytically
\cite{EFE}. As already discussed, in this paper we improve the
Newtonian modelling by using an effective gravitational potential in
order to mimic relativistic effects in the framework of Newtonian
hydrodynamics: equation (\ref{def:hatV}) is where this potential must
step in. In place of the Newtonian potential, following
\cite{RamppJanka2002}-\cite{Dimmelmeier2006} who made the same
assumption in hydrodynamical simulations of stellar core collapse and
post-bounce evolution, we use the potential $\Phi_{TOV}$ which is
solution of the equations of hydrostatic equilibrium in general
relativity:
\begin{eqnarray}
\label{def:dPhiTOVdr}
\frac{d\Phi_{TOV}}{dr}&=&\frac{[\eps(r)+P(r)][m_{TOV}(r)+4\pi r^3P(r)]}{\rho(r)
r[r-2m_{TOV}(r)]}\\
\label{def:mTOV}
m_{TOV}(r) &=& 4\pi\int_0^r dr'r'^2\eps(r').
\end{eqnarray}
Our goal is therefore achieved by plugging $d\Phi_{TOV}/dr$ in
(\ref{def:hatV}) in place of $d\Phi/dr$.  Further details about this
approximation are given in \ref{app:pseudoTOV}.

In Eqs. (\ref{def:dPhiTOVdr}), (\ref{def:mTOV}) $M_{TOV}=m_{TOV}(R)$
is the gravitational (or TOV) mass of the NS, and the pressure ($P$),
rest-mass density ($\rho$) and energy density ($\eps$) profiles are
determined from the relativistic Tolman-Oppenheimer-Volkoff stellar
structure equations.  For the sake of simplicity, in these formulas we
have omitted all carets on the equilibrium quantities.

\subsection{The Dynamics Equations}\label{ssec:dyn}
The Lagrangian (\ref{Linner}) for the star internal dynamics can
easily be found by collecting the terms given in Eqs.\,(\ref{L_T}),
(\ref{Iij}), (\ref{Lbody}), (\ref{T_I=sum}), (\ref{def:U}) and
(\ref{def:V}); the Lagrange equations for the five fluid variables
$q_i=\{a_1,a_2,a_3,\phi,\lambda\}$ yield
\begin{eqnarray}
\label{dda1}
\ddot a_1 &=& a_1(\Lambda^2+\Omega^2) - 2a_2\Lambda\Omega
+ \frac{1}{2}\frac{\hat{V}}{\wh{\mathcal{M}}}R_{NS}^3a_1\tilde{A}_1
+ \frac{R_{NS}^2}{\wh{\mathcal{M}}}\frac{\Pi}{a_1} - c_{11}a_1\\
\nonumber
\ddot a_2 &=& a_2(\Lambda^2+\Omega^2) - 2a_1\Lambda\Omega
+ \frac{1}{2}\frac{\hat{V}}{\wh{\mathcal{M}}}R_{NS}^3a_2\tilde{A}_2
+ \frac{R_{NS}^2}{\wh{\mathcal{M}}}\frac{\Pi}{a_2} - c_{22}a_2\\
\nonumber
\ddot a_3 &=& \frac{1}{2}\frac{\hat{V}}{\wh{\mathcal{M}}}R_{NS}^3a_3
\tilde{A}_3 + \frac{R_{NS}^2}{\wh{\mathcal{M}}}\frac{\Pi}{a_3} - c_{33}a_3\\
\nonumber
\dot{J}_s &=& \frac{\wh{\mathcal{M}}}{R_{NS}^2}c_{12}(a_2^2-a_1^2)\\
\nonumber
\dot{\mathcal{C}} &=& 0
\nonumber
\end{eqnarray}
where $\Pi$ is given by (\ref{def:Pi}) and where we have defined
\[
\tilde{A}_i = \int_0^\infty\frac{du}{\Delta(u)(a_i^2+u)}
\]
and
\begin{eqnarray}
\label{def:J_s}
J_s = \frac{\wh{\mathcal{M}}}{R_{NS}^2}[(a_1^2+a_2^2)\Omega-2a_1a_2
\Lambda],\quad
\mathcal{C} = \frac{\wh{\mathcal{M}}}{R_{NS}^2}[(a_1^2+a_2^2)\Lambda-2a_1a_2
\Omega].
\end{eqnarray}
$J_s$ and $\mathcal{C}$ are, respectively, the spin angular momentum
of the star, and a quantity proportional to its circulation in the
locally nonrotating inertial frame. Notice that $\dot{\mathcal{C}}=0$
because we work in absence of viscosity. In this paper we consider
irrotational models, i.e. with $\mathcal{C}=0$.

Equations (\ref{dda1}) are a generalised version of Eqs.\,(31)-(35) in
\cite{WL2000}; they are more general in the sense that (1) they are
\emph{not} restricted to the use of a polytropic EOS but are valid for
any barotropic EOS and (2) they are written for any scalar
gravitational potential for the self-gravity of the NS. When testing
our programmes in order to reproduce the results of \cite{WL2000}, we
shall adopt the Newtonian potential in (\ref{def:hatV}),
i.e. $d\Phi/dr=4\pi\rho r$; elsewhere we will make the choice of using
the effective TOV potential (\ref{def:dPhiTOVdr})-(\ref{def:mTOV}).

We now reduce the equations (\ref{dda1}) to coupled algebraic
equations by demanding quasi-equilibrium: during its evolution, in
fact, a coalescing BH-NS binary will likely follow a quasi-equilibrium
sequence with constant circulation (in particular, we set
$\mathcal{C}=0$). Physically this statement relies on the fact that in
such binaries the ratio between the orbital decay time due to gravity
waves ($t_{gw}$) and the tidal synchronization time ($t_{syn}$),
i.e. the quantity governing the relative importance of viscosity, is
smaller than unity \cite{NoSync}. This allows one to consider the
fluid body not to be tidally locked with the orbital motion: internal
fluid motion is therefore a necessary ingredient of the model. For a
binary system in quasi-equilibrium one has to require that
\begin{eqnarray}
\label{QE1}
\ddot{a}_i = 0&&\\\nonumber
\phi = \Psi&&\\\nonumber
\dot{\phi} = \Omega = \dot{\Psi} = \sqrt{\frac{M_{BH}}{r^3}}.&&\nonumber
\end{eqnarray}
These are the equations we solve in this paper; in order to do so we
adopt a Newton-Raphson scheme. The quasi-equilibrium sequence with
constant circulation is parametrized by the binary orbital separation
$r$. What we have is therefore essentially a ``hydro without hydro''
method, that is, in order to include finite-size effects in the
inspiral phase we do not solve the hydrodynamic equations explicitly,
but instead use ``snapshots'' generated by quasi-equilibrium
conditions. To obtain a quasi-equilibrium sequence, we start by
placing a non rotating spherical star in equilibrium at a distance
$r_0\gg R_{NS}$ from the black hole\footnote{We make sure that the
sequence we obtain is independent of $r_0$.}, we then gradually reduce
the orbital separation, solve (\ref{QE1}) and monitor the stars axes
until a critical separation $r_{tide}$ is reached, at which no
quasi-equilibrium configuration is possible. This critical distance is
identified exploiting the fact that the algorithm cannot find any
solution to the system (\ref{QE1}) or by keeping track of the
numerically calculated derivative $\partial
r_{norm}/\partial(a_2/a_1)$, where
\begin{eqnarray}
r_{norm}=\frac{r}{R_{NS}}\left(\frac{M_{NS}}{M_{BH}}\right)^{1/3},
\end{eqnarray}
which tends to zero at tidal disruption: both definitions yield the
same values of $r_{tide}$\footnote{An alternative approach to evaluate 
$r_{tide}$, based on the estimate of the Roche lobe radius in
a Newtonian framework, has been used in \cite{LP07}.}.

\section{Results}\label{sec:results}
In this section we firstly compare our results on the tidal disruption
of a TOV star to data extracted from the Full-GR results of Taniguchi
et al.\,\cite{TBFS08}: this allows us to assess the range of validity
of our model and to prove the improvement gained over calculations
performed with a Newtonian ellipsoidal star. Subsequently, we show and
discuss some examples of quasi-equilibrium sequences in the case of a
$1.4\,M_\odot$ NS in tidal interaction with several kinds of stellar
mass BHs; for the NS EOS we take two cases\footnote{See
\ref{ssec:NewRes} for details.}: (1) the (stiff) EOS APR2, which
yields $R_{NS}=11.53\,$km, and the (soft) EOS BALBN1H1, which instead
gives $R_{NS}=12.84\,$km.

\subsection{Comparison with Previous Results}
As a very first step we calculated BH-NS quasi-equilibrium sequences
choosing a Newtonian gravitational potential for the NS, and a
polytropic EOS with $n=1$. We successfully reproduced the tables for
$r_{tide}$ given in Wiggins and Lai's work \cite{WL2000} and their
data on equilibrium sequences.

\begin{figure}[!ht]
\begin{center}
\includegraphics[scale=.23,angle=-90]{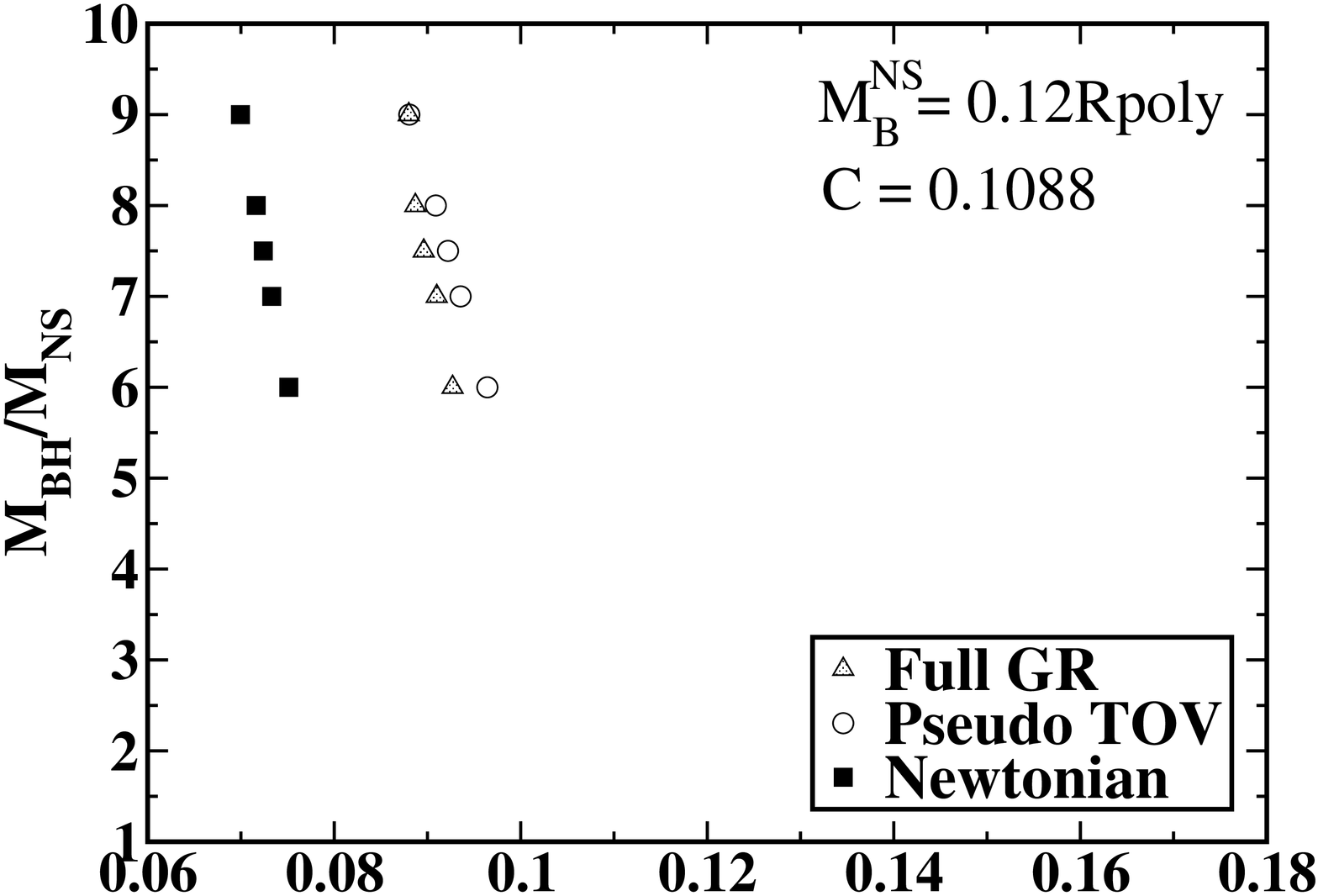}
\includegraphics[scale=.23,angle=-90]{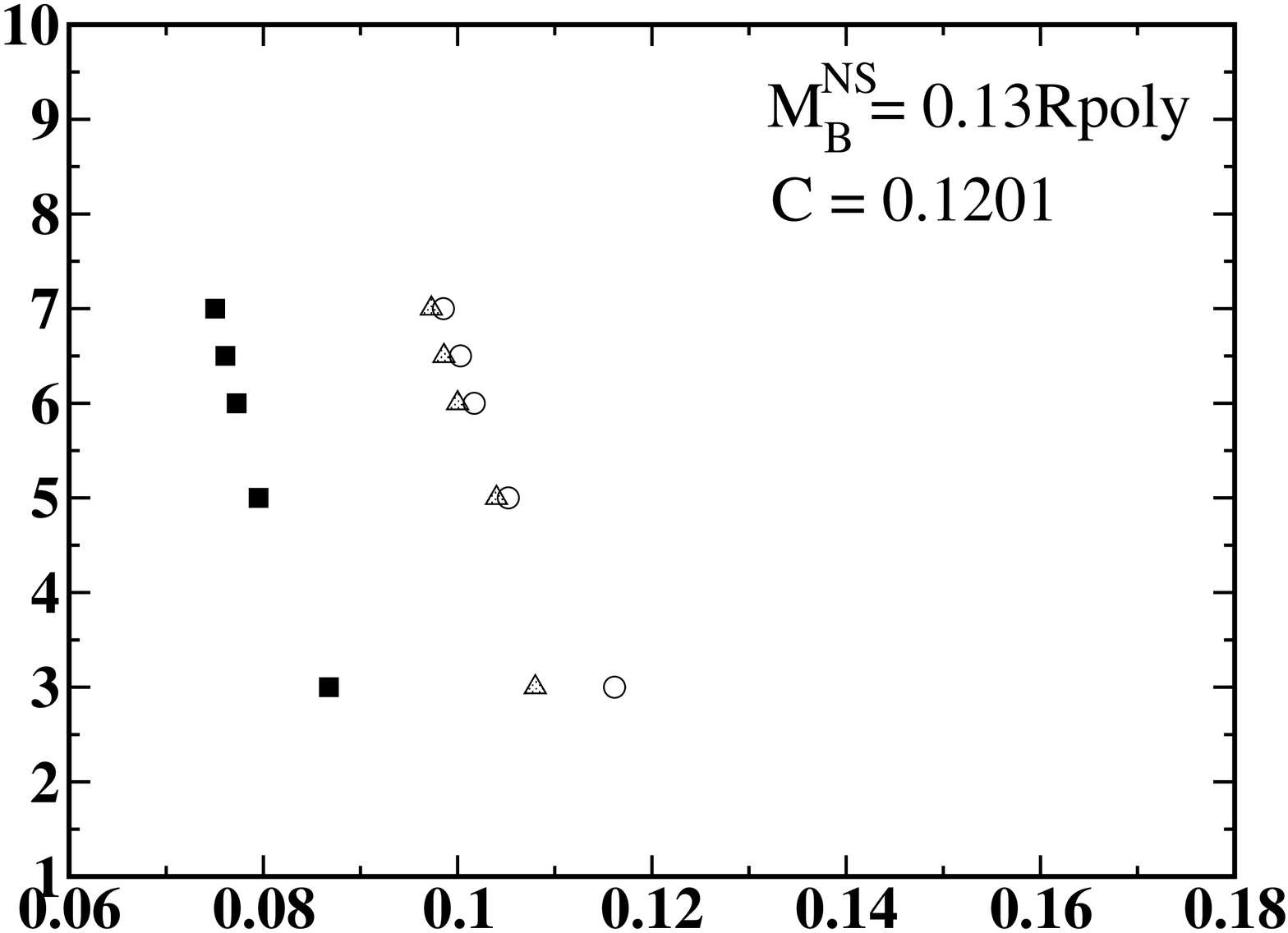}
\includegraphics[scale=.23,angle=-90]{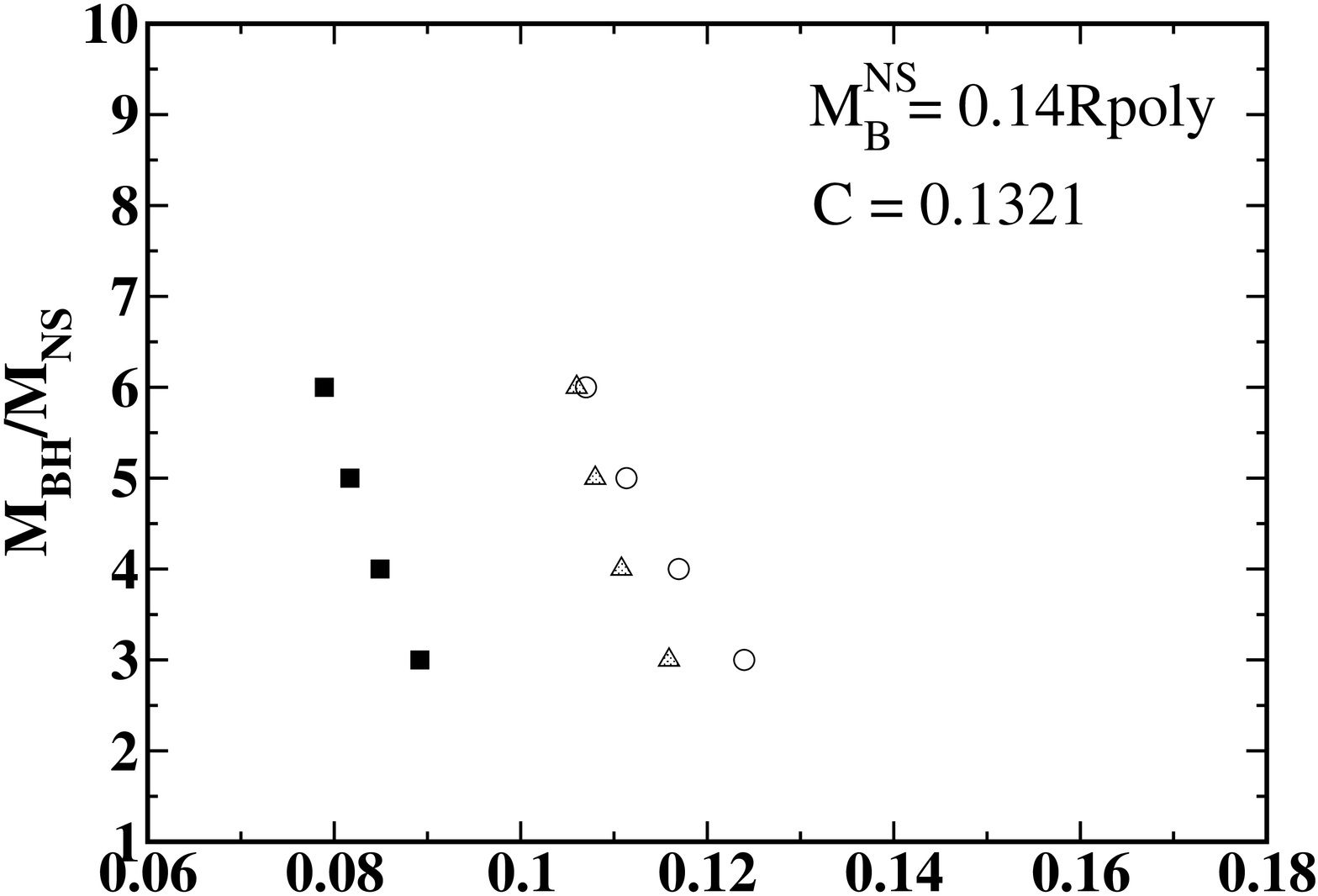}
\includegraphics[scale=.23,angle=-90]{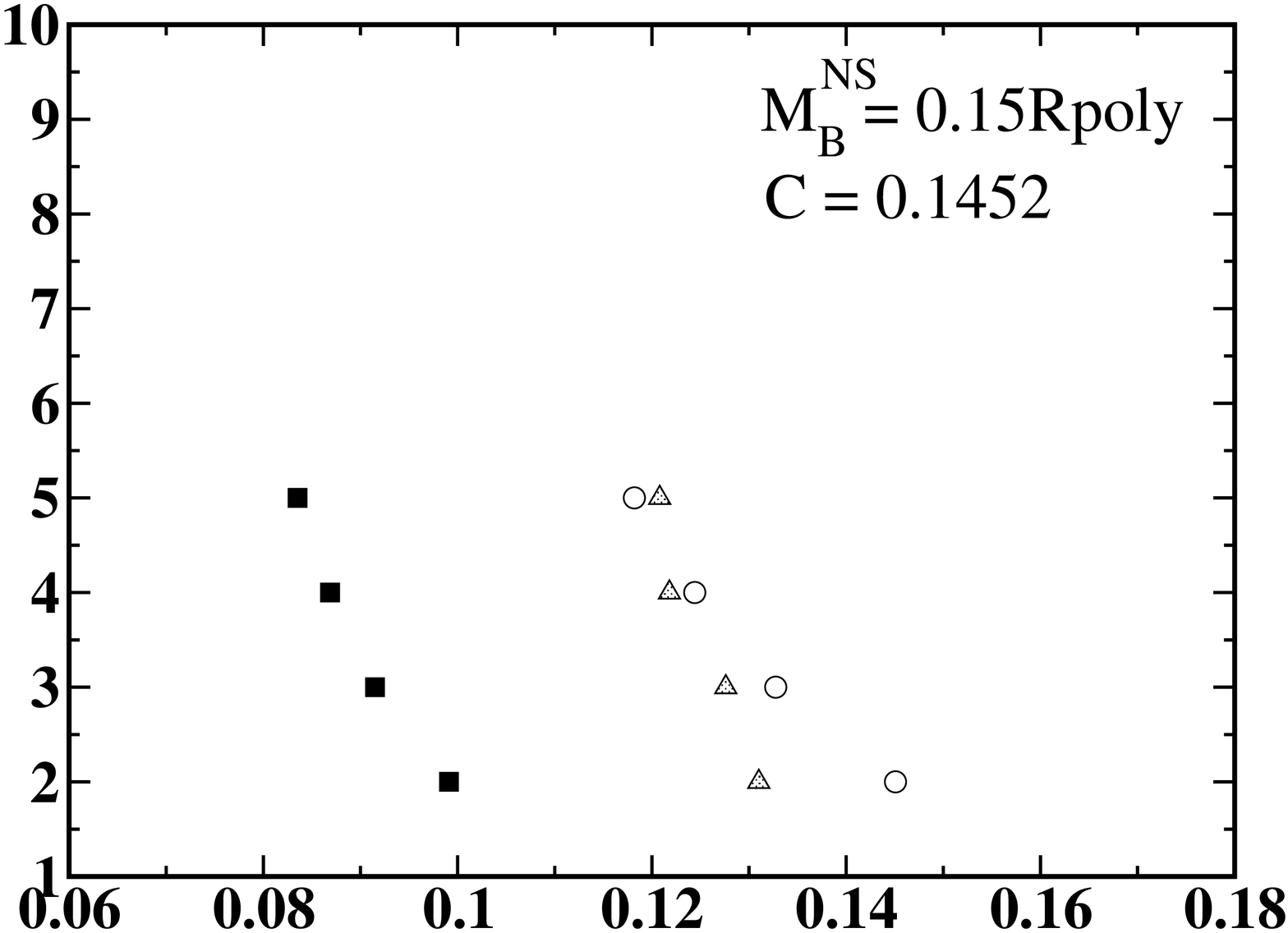}
\includegraphics[scale=.23,angle=-90]{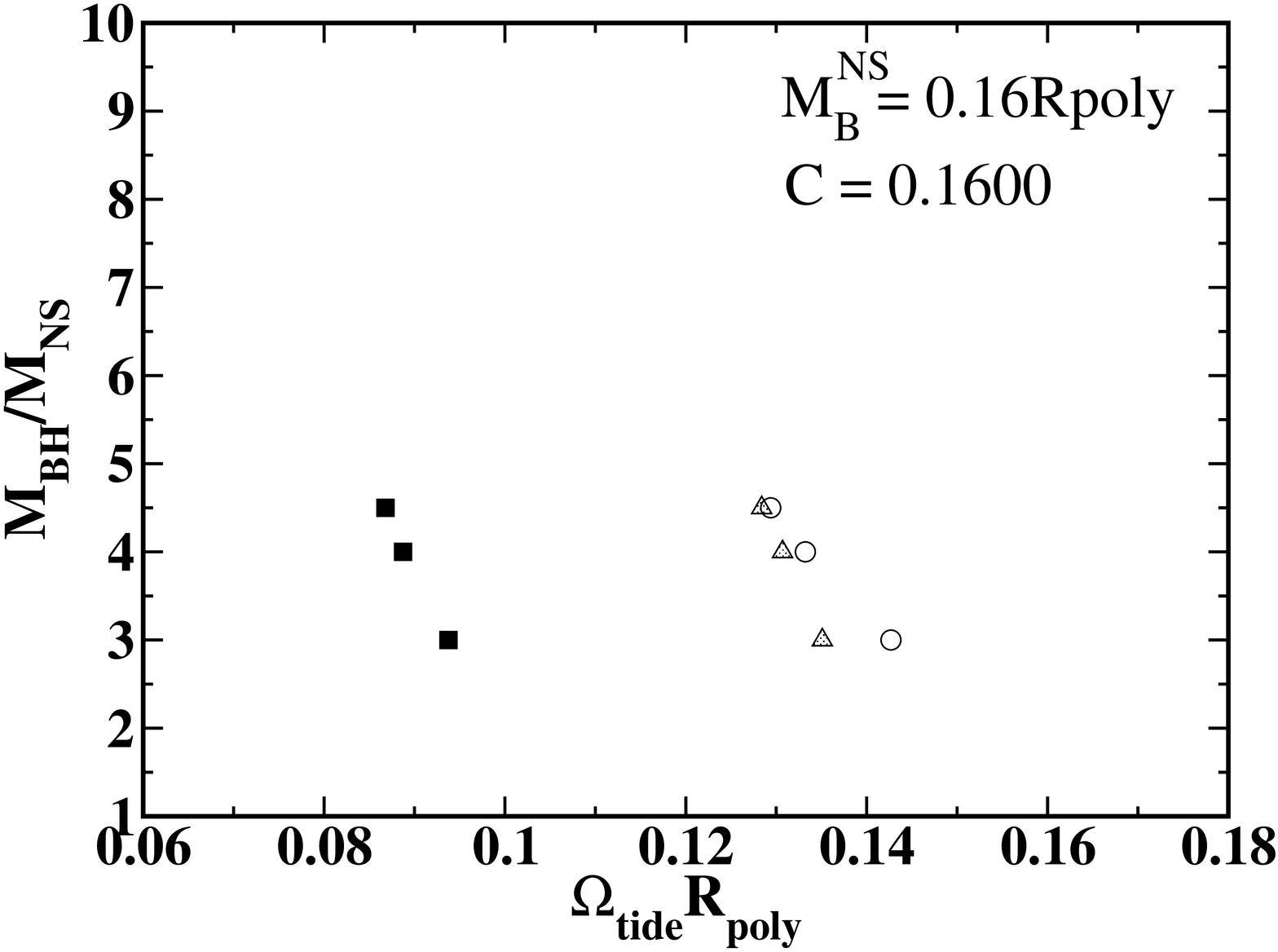}
\includegraphics[scale=.23,angle=-90]{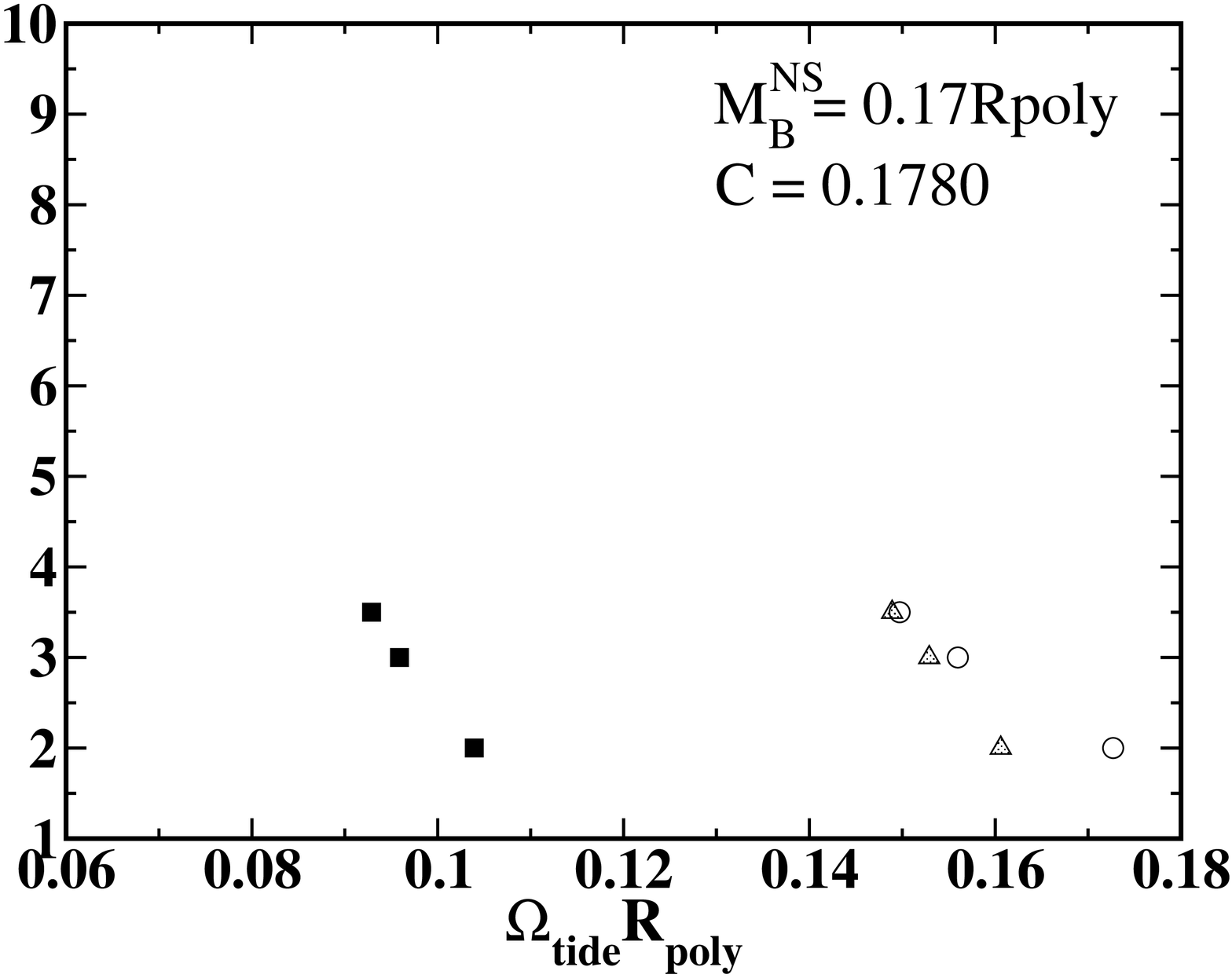}
\end{center}
\caption{Comparison between Full-GR results from \cite{TBFS08}
  (Full-GR), results obtained with the Newtonian ellipsoidal model of
  \cite{WL2000} (Newtonian) and results from our improved ellipsoidal
  model (Pseudo TOV).  Each graph shows the mass ratio
  (Eq.\,(\ref{def:q})) versus the tidal disruption limit calculated as
  in Eq.\,(\ref{OmTide}).  In all cases the neutron star EOS is an
  $n=1, \kappa=1$ polytrope.  The compactness $C$ indicated in each
  panel refers to the Full-GR and to our models.  }
\label{TBFS}
\end{figure}

After these preliminary tests, we move on to using the improved
gravitational potential for the NS and compare results given by our
model with the Full-GR results presented in Figure 11 of \cite{TBFS08}
by Taniguchi, Baumgarte, Faber and Shapiro, who study the tidal
disruption of $n=1$ and $\kappa =1$ polytropic NSs, orbiting
Schwarzschild BHs on equatorial circular orbits. In Figure \ref{TBFS}
we show the results of this comparison; to facilitate the reader in
analysing this figure along with Figure 11 of \cite{TBFS08}, we use
the same physical quantities as Taniguchi et al. and hence the mass
ratio
\begin{eqnarray}
\label{def:q}
q=\frac{M_{BH}}{M_{NS}}
\end{eqnarray}
is plotted as a function of
\begin{eqnarray}
\label{OmTide}
\Omega_{tide}R_{poly} = \sqrt{\frac{M_{BH}+M_{NS}}{r_{tide}^3}}
\sqrt{\kappa^n},
\end{eqnarray}
which is the orbital angular velocity at which the tidal disruption
occurs, normalised with respect to the polytropic length scale
$R_{poly}=\sqrt{\kappa^n}$; in these definitions, $M_{BH}$ is the hole
mass (which is fixed by the value of $q$) and $M_{NS}$ is the ADM mass
of the isolated star, which coincides with the gravitational-TOV mass.
Of course, we take the same TOV equilibrium models used in
\cite{TBFS08}, whose properties are discussed in the tables at the end
of the aforementioned paper; the neutron star mass is fixed by
choosing a specific value for the baryonic mass
\begin{eqnarray}
M_B^{NS} = 4\pi \int_0^{R_{NS}}
\frac{\hat{r}^2\hat{\rho}(\hat{r})}{\sqrt{1-2\hat{m}_{TOV}(\hat{r})/
    \hat{r}}}d\hat{r},
\end{eqnarray}
where $\hat{m}_{TOV}(\hat{r})$ is defined by (\ref{def:mTOV}).  In
Figure \ref{TBFS} we also show the results yielded by Wiggins and
Lai's original approach: in this case the star model is fixed by
equating the Newtonian star mass $M_N$ to $M_{NS}$.  In each panel of
the figure, the value of the stellar compactness $C=M_{NS}/R_{NS}$ for
the relativistic configurations is also displayed.

The results displayed in Figure \ref{TBFS} show that:
\begin{itemize}
\item our pseudo-relativistic ellipsoidal model agrees with relativistic data
  much better than the Newtonian ellipsoidal model;
\item for a given compactness, the agreement between our data and
  Full-GR data improves as the mass ratio increases. This is due to
  the fact that our model assumes that the centre of mass of the star
  moves on a geodesic of the black hole spacetime in a reference frame
  centered on the BH, a condition which is better satisfied for larger
  mass ratios.  For instance, for $q=9$ and $M_B^{NS}=0.12R_{poly}$
  ($C=0.1088$), the results practically coincide;
\item for lower values of $q$ the agreement between our data and the Full-GR
  data increases as the stellar compactness increases.
\end{itemize}
For a more detailed and quantitative comparison, in \ref{app:table} we
tabulate the values of $\Omega_{tide}R_{poly}$ plotted in Figure
\ref{TBFS}.

\clearpage
\subsection{BH-NS Equilibrium Sequences and Tidal Disruption Limits}
\label{ssec:NewRes}
Having tested the validity of our approach, we employ it to study
different possible BH-NS binary configurations, determining the
quasi-equilibrium sequences and the tidal disruption radius $r_{tide}$
for equatorial circular orbits. We describe the fluid forming the NS
with two different EOS proposed in recent years by the nuclear physics
community, which we call APR2 and BALBN1H1.
\begin{itemize}
\item The Akmal-Pandharipande-Ravenhall (APR2) hadronic EOS
  \cite{APR2} describes matter consisting of neutrons, protons,
  electrons and muons in weak equilibrium; it is obtained within
  nuclear many-body theory using a variational approach to the
  Shr\"{o}dinger equation; its microscopic input is based on the
  Argonne $v_{18}$ potential for nucleon-nucleon interactions
  \cite{Argonnev18} --- which is calibrated to deuteron properties and
  vacuum nucleon-nucleon phase shifts for laboratory energies
  $E_{lab}$ up to $350\,$MeV --- and on the Urbana IX ($UIX$)
  three-body potential \cite{UrbanaIX}; relativistic corrections to
  \emph{both} potentials are included \cite{RelHams} (which yields
  $v_{18}+\delta v+UIX^*$)
\item The Balberg-Gal (BALBN1H1) EOS \cite{BALBN1H1} describes matter
  consisting of neutrons, protons, electrons, muons and hyperons
  ($\Sigma$, $\Lambda$ and $\Xi$) in equilibrium. Assuming the mean
  field approximation, an effective potential is employed, whose
  parameters are tuned in order to reproduce the properties of nuclei
  and hypernuclei according to high energy experiments.  This EOS is a
  generalization of the Lattimer-Swesty EOS, which does not include
  hyperons.
\end{itemize}
APR2 is much stiffer than BALBN1H1 and, in terms of stiffness, most
modern EOS fall in between the two equations of state we choose.

We consider several values of the mass ratio $q=M_{BH}/M_{NS}$,
ranging from $5$ to $50$, and three different values of the black hole
spin: $a=0$, $a=0.5M_{BH}$ and $a=0.99M_{BH}$.

\begin{figure}[!ht]
\begin{center}
\includegraphics[scale=.23,angle=-90]{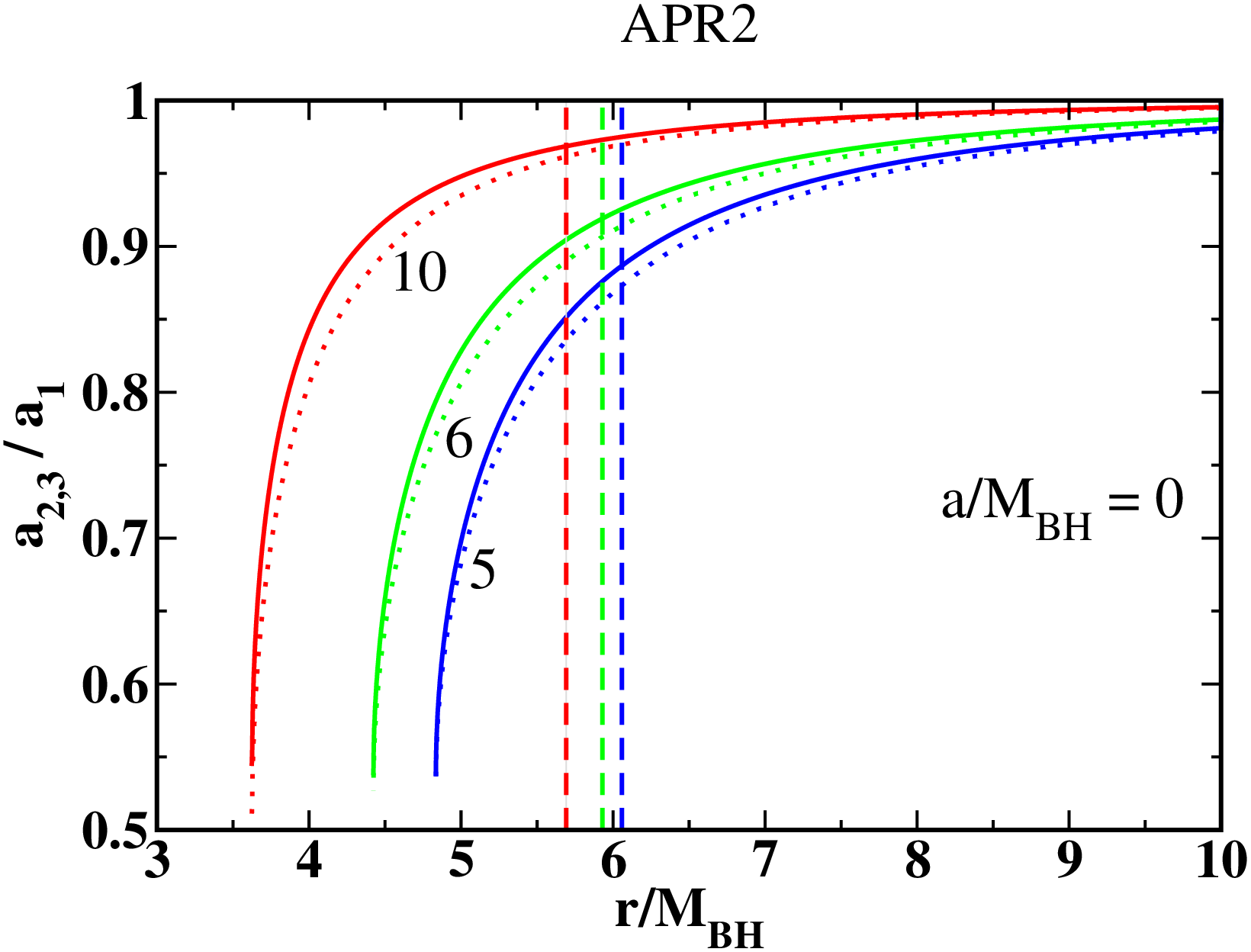}
\includegraphics[scale=.23,angle=-90]{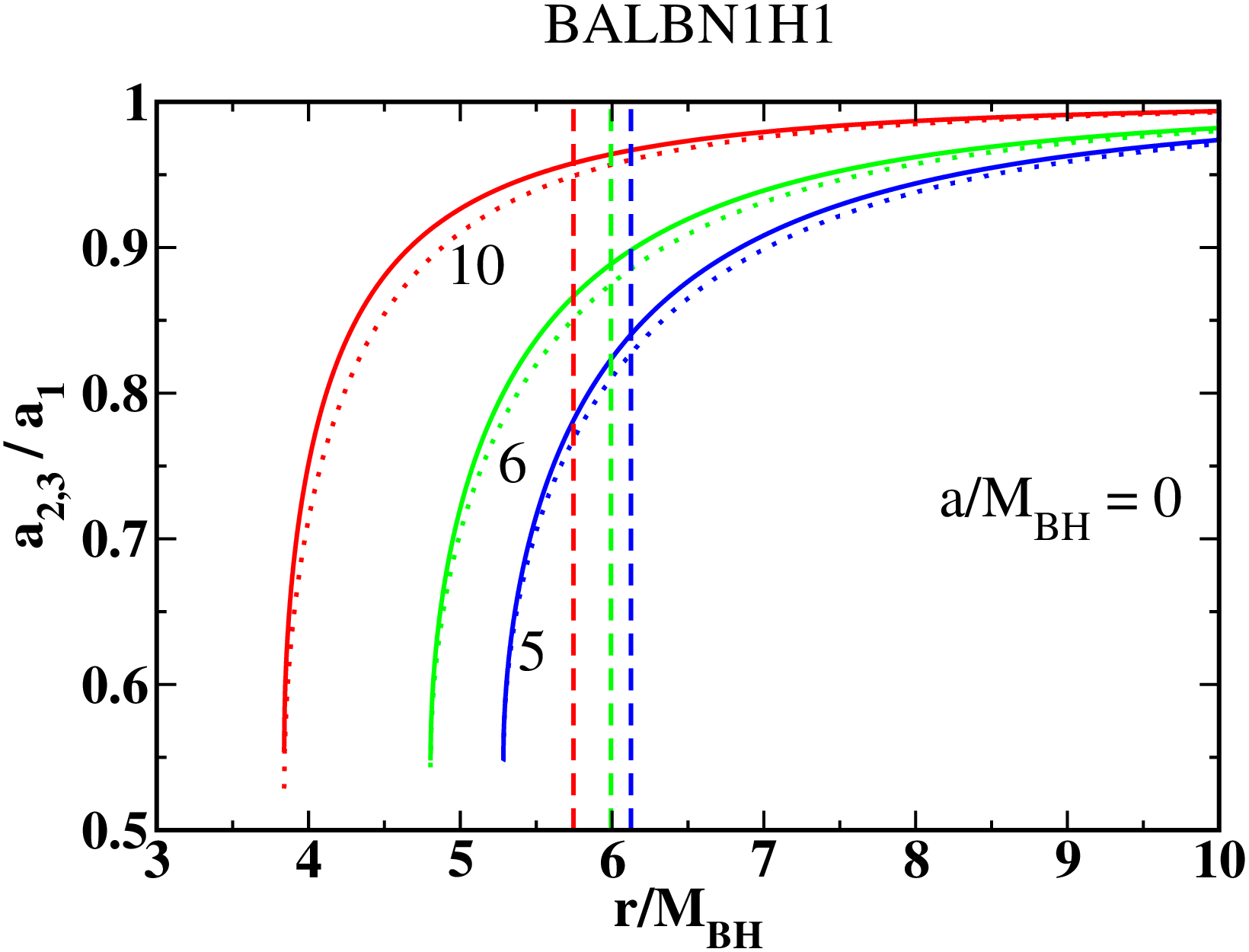}
\includegraphics[scale=.23,angle=-90]{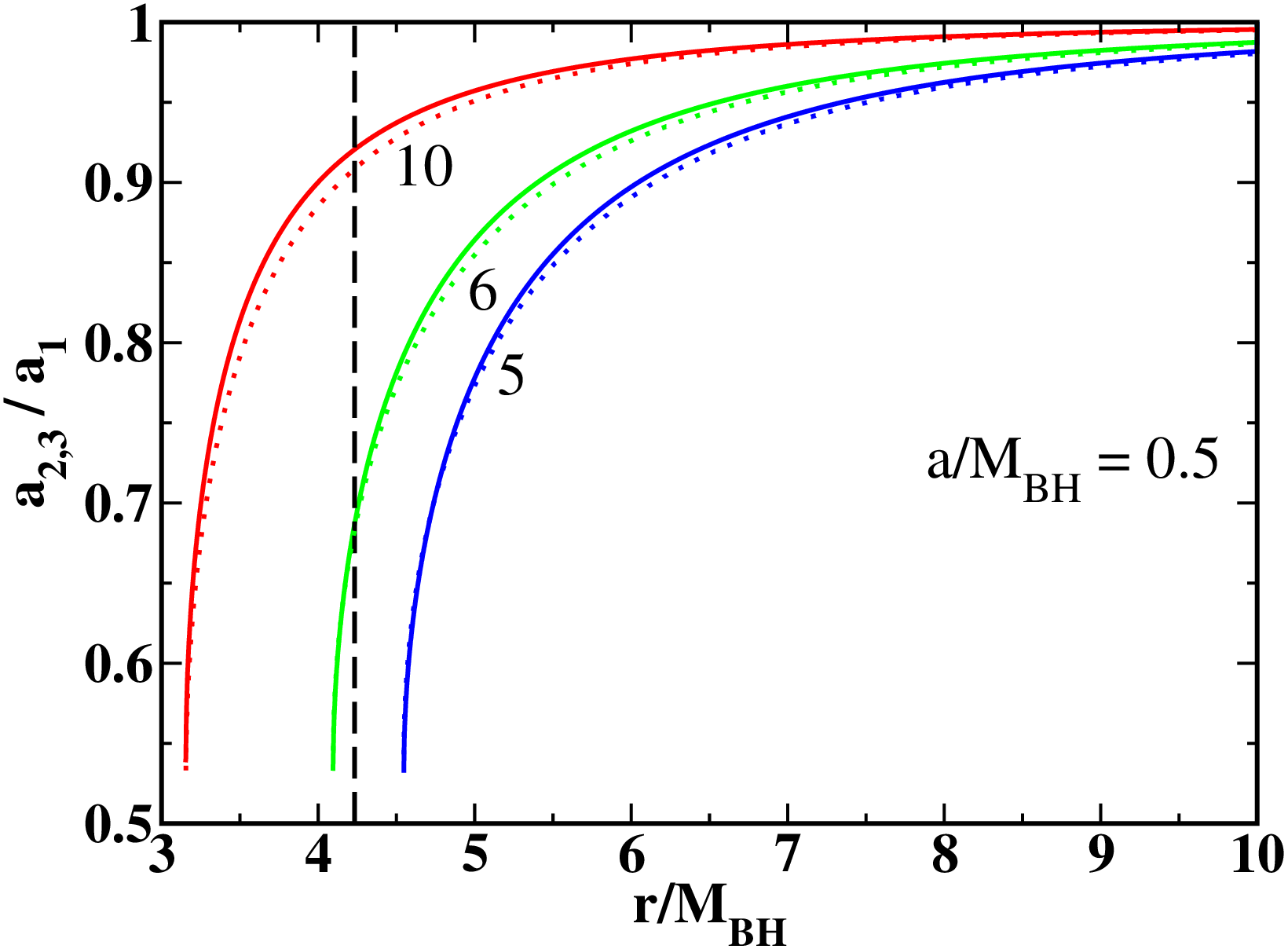}
\includegraphics[scale=.23,angle=-90]{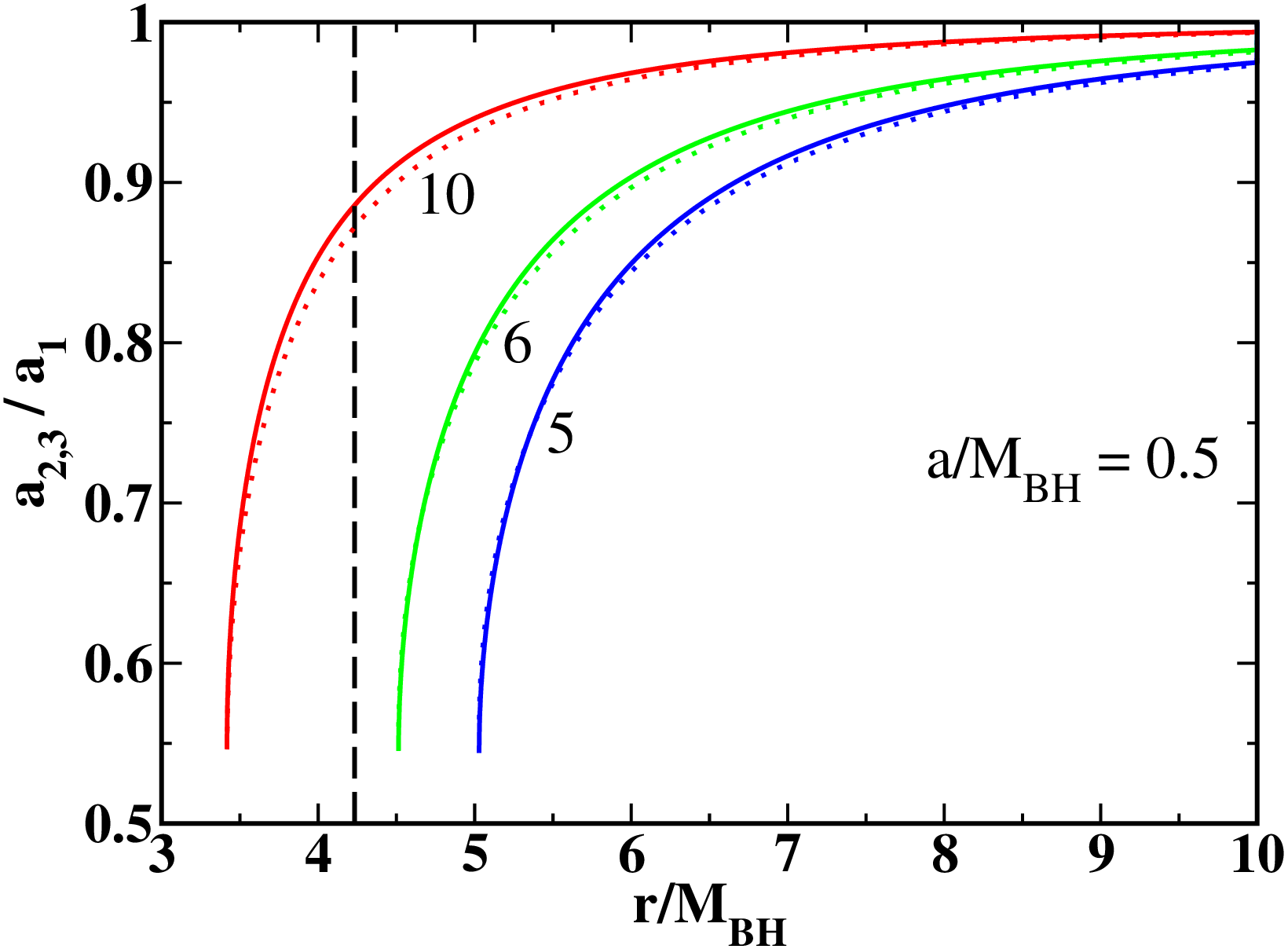}
\includegraphics[scale=.23,angle=-90]{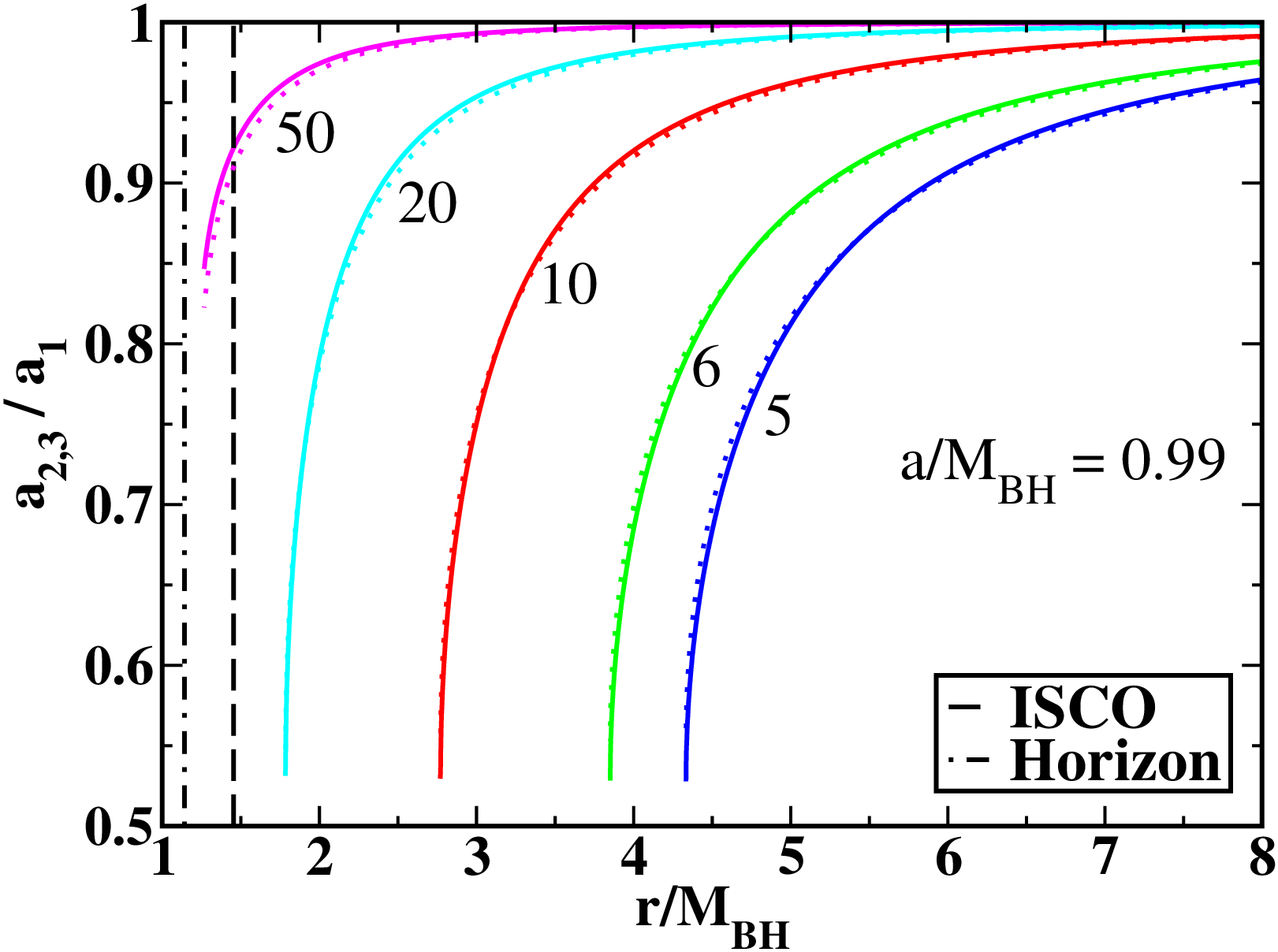}
\includegraphics[scale=.23,angle=-90]{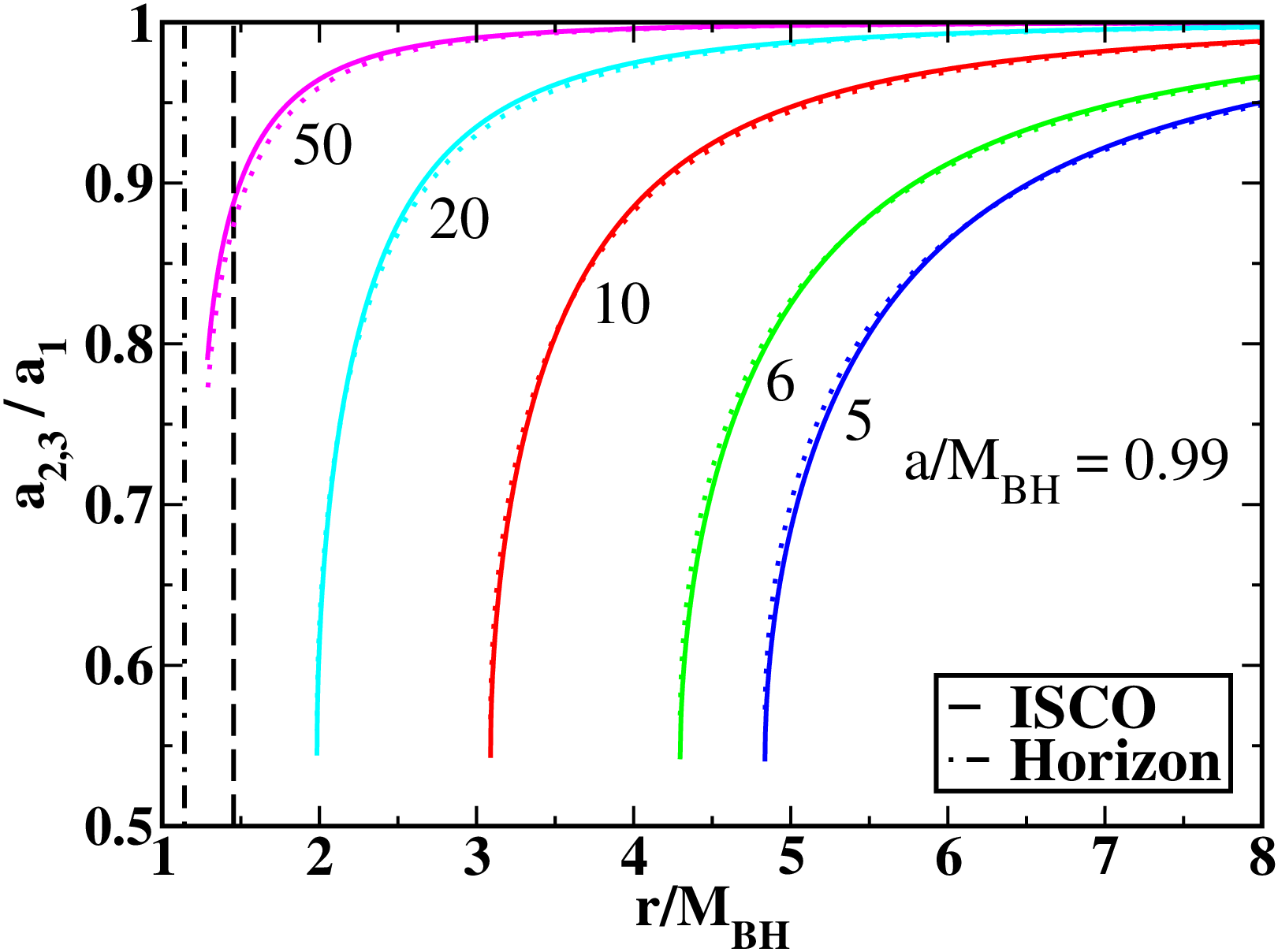}
\end{center}
\caption{$a_2/a_1$ (continuous lines) and $a_3/a_1$ (dotted lines) are
  plotted as functions of $r/M_{BH}$ for a NS orbiting a black
  hole. The angular momentum of the BH takes the values
  $a/M_{BH}=\{0,0.5,0.99\}$ as indicated in each panel. The mass ratio
  $q=M_{BH}/M_{NS}$ is indicated by the numbers next to each curve,
  which represent a quasi-equilibrium sequence. $M_{NS}=1.4\,M_\odot$
  in all graphs: in the left column the star is modeled using APR2 EOS
  ($R_{NS}= 12.84\,$km), in the right column using BALBN1H1 EOS
  ($R_{NS}=11.53\,$km).  The ISCOs are shown by the dashed vertical
  lines, while the dot-dashed lines in the lower graphs mark the
  position of the black hole horizon. In the $q=50$ case, the sequence
  does not terminate with tidal disruption; it stops when
  $r=a_1+R_{BH}^+$, where $R_{BH}^+$ is the size of the black hole
  outer horizon at the equatorial plane.}
\label{6grafici}
\end{figure}

In order to understand which sequences are possible candidates to be
the engine of a SGRB, we compare the tidal disruption orbital
separations $r_{tide}$ with $r_{ISCO}$.  In the Schwarzschild case, we
determine the location of the ISCO through the analytic fit given in
\cite{TBFS08}, where the effects of the finite mass ratio and of the
stellar compactness are taken into account:
\begin{eqnarray}
r_{ISCO} = M_{BH}\frac{1+q^{-1}}{6^{-3/2}\left( 1-0.444q^{-1/4}(1-3.54C^{1/3})
  \right)}.
\end{eqnarray}
Note that according to this equation the values of $r_{ISCO}/M_{BH}$
we give for $a=0$ in Figure \ref{6grafici} depend on the mass ratio
$q$.

Since no such fit exists for the Kerr BH-NS binaries, when $a\neq0$ we
estimate the ISCO by using the formulae derived in \cite{BPT1972} for
a point mass in the gravitational field of a Kerr BH:
\begin{eqnarray}
r_{ISCO} = M_{BH}\{3+Z_2\mp [(3-Z_1)(3+Z_1+2Z_2)]^{1/2}\}&&\\
\nonumber
Z_1 = 1+(1-a^2/M_{BH}^2)^{1/3}[(1+a/M_{BH})^{1/3}+(1-a/M_{BH})^{1/3}] &&\\
\nonumber
Z_2 = (3a^2/M_{BH}^2+Z_1^2)^{1/2},
\end{eqnarray}
where the upper sign holds for corotating orbits and the lower sign
for counterotating orbits. Note that according to these equations the
values that $r_{ISCO}/M_{BH}$ assumes for $a\neq0$ do not depend on
$q$.

The results of our numerical integrations are displayed in Figure
\ref{6grafici}, where the ratios $a_2/a_1$ (continuous lines) and
$a_3/a_1$ (dotted lines) among the NS axes are shown as functions of
$r/M_{BH}$; the quasi-equilibrium sequences end when the tidal
disruption of the NS is reached (see \S \ref{ssec:dyn}). The left
panels refer to the EOS APR2, the right ones to BALBN1H1; the upper
panels refer to the Schwarzschild case, the middle panels to
$a=0.5M_{BH}$, the lower panels to $a=0.99M_{BH}$. The dashed,
vertical lines are the locations of the ISCO. In the $a=0.99M_{BH}$
case we also indicate the location of the black hole outer horizon
$R_{BH}^+$ with a dot-dashed vertical line. From the graphs in Figure
\ref{6grafici} we extract the following information.
\begin{itemize}
\item In each panel, the larger the mass ratio $q$ is, the closer the
  NS can get to the BH before being disrupted, i.e. $r_{tide}/M_{BH}$
  decreases as $q$ increases. Therefore the conditions of the SGRB
  mechanism are more likely to be satisfied for low values of $q$.
\item If $a=0$ (first row) the star enters the ISCO before being
  disrupted for all the considered values of $q$, and for both the
  APR2 and the BALBN1H1 EOS: therefore, SGRBs cannot be ignited. On
  the other hand, if $a=0.5M_{BH}$ (middle row) the star is disrupted
  at $r>r_{ISCO}$ for $q\lesssim5.7$ (APR2 EOS) and $q\lesssim6.6$
  (BALBN1H1 EOS) and the SGRB may take place. If, finally,
  $a=0.99M_{BH}$ (third row), the star is disrupted at $r>r_{ISCO}$
  for $q\lesssim28$ (APR2 EOS) and $q\lesssim33$ (BALBN1H1 EOS) and,
  again, the SGRB may take place.

  We note that as the the black hole spin increases, both
  $r_{tide}/M_{BH}$ and $r_{ISCO}/M_{BH}$ decrease; however,
  $r_{ISCO}/M_{BH}$ decreases more rapidly than $r_{tide}/M_{BH}$, and
  consequently for a given mass ratio $q$, chances to develop an SGRB
  are higher if the black holes rotates faster.  Thus, in general the
  conditions for the SGRB mechanism proposed in \cite{GRB1} to take
  place are favoured for low values of $q$ and high values of $a$.
\item Comparing the two graphs in each row, we see that for NSs with a
  stiffer EOS, $r_{tide}/M_{BH}$ is smaller. This is what one expects,
  since a star with a stiffer EOS is more compact and thus less
  deformable: hence a stronger gravitational field is needed to
  disrupt the star.
\item As the value of the BH spin increases, the values of the axes
  $a_2$ (continuous lines) and $a_3$ (dotted lines) tend to coincide.
\end{itemize}

We also note that, if $a=0.99M_{BH}$, for $q\simeq 20$ the star enters
the ergosphere ($r=2M$, since $\theta=\pi/2$) before disruption.

\clearpage
\section{Conclusions}\label{sec:conclusions}
In this paper we have studied the effects of a Kerr BH tidal field on
a NS; we have focused on determining the orbital separation at which
the NS is tidally disrupted ($r_{tide}$) and the star
quasi-equilibrium sequence for several BH-NS binaries.  Our work is
based on the affine model, which we have improved with respect to
previous works, by describing the NS self-gravity with an effective
relativistic scalar potential (Eqs.\,(\ref{def:dPhiTOVdr}) and
(\ref{def:mTOV})), and by using more realistic equations of state for
the NS matter.  This approach has the advantage of allowing a quick
computation of $r_{tide}$ and of the quasi-equilibrium sequences for
any value of the binary mass ratio $q$ and of the BH spin parameter
$a$.

A comparison of the results obtained using our approach for NS
disruption with Full-GR results, which are available in the case of
non-rotating BHs and polytropic NSs, shows an excellent agreement up
to small values of $q$ (which depend on the NS compactness), beyond
which our model starts to \emph{underestimate} $r_{tide}$; this
happens because we assume that the centre of mass of the star moves on
along a BH geodesic, a condition which is better satisfied for larger
mass ratios.  Our model agrees with relativistic calculations much
better than ellipsoidal models in which the neutron star self-gravity
is treated at a Newtonian level.

Given the very good results of this test, we have determined the
quasi-equilibrium sequences and the tidal disruption radii of several
BH-NS binaries in order to evaluate with our model the role played by
(1) the BH spin parameter $a$ and (2) the equation of state of the NS
fluid.  We found that the soft gamma-ray burst scenario is favoured by
\emph{low} values of the mass ratio $q$ and \emph{high} values of the
BH spin parameter $a$.  In addition, the higher the BH spin parameter
is, the more the values of the NS axes $a_2$ and $a_3$ (i.e. the
principal axes which do not point at the BH) tend to coincide.

In quantitative terms we have shown that, for the stiffer equation of
state APR2 a SGRB may occur if $q\lesssim 28$ when $a=0.99\,M_{BH}$ or
if $q\lesssim 5.7$ when $a=0.5\,M_{BH}$.  For the softer equation of
state BALBN1H1 a SGRB may develop if $q\lesssim 33$ when
$a=0.99\,M_{BH}$ or if $q\lesssim 6.6$ when $a=0.5\,M_{BH}$.

As a future development of our work, we intend to study the dynamical
behaviour of BH-NS binaries and to determine the gravitational
radiation emitted in the late phases of their inspiral. This means
taking into account tidal interaction corrections to the orbital
dynamics and to the gravitational waveform, and thus will require the
inclusion in our model of a precise treatment of the orbit. Moreover
we would like to improve the results of our model for low values of
the mass ratio $q$.

\appendix
\section{Effective General Relativistic Gravitational
  Potential}\label{app:pseudoTOV}
In a Newtonian context, the gravitational potential of a non-rotating
star at equilibrium is governed by the Poisson equation and the
hydrostatic equilibrium equation, i.e.
\begin{eqnarray}
\label{Newt}
\Delta\Phi_{Newt}&=&4\pi\rho\\
\nonumber
\frac{d\Phi_{Newt}}{dr} &=& -\frac{1}{\rho}\frac{dP}{dr},
\end{eqnarray}
and the  system is closed by choosing an equation of state.

In General Relativity, the equilibrium configuration of a spherical
distribution of fluid is determined by the Tolman-Oppenheimer-Volkoff
(TOV) equations of hydrostatic equilibrium
\begin{eqnarray}
\label{TOV}
\frac{dm(r)}{dr}&=&4\pi r^2\epsilon(r)\\
\nonumber
\frac{d\nu(r)}{dr}&=&2\frac{m(r)+4\pi r^3P(r)}{r(r-2m(r))}\\
\nonumber
\frac{dP(r)}{dr}&=&-\frac{(P(r)+\epsilon(r))}{2}\frac{\nu(r)}{dr}
\end{eqnarray}
where $m(r)$ is the gravitational mass enclosed within a radius $r$;
also this system must be closed by choosing an equation of state.  In
order to define the \emph{effective relativistic} or
\emph{pseudo-relativistic scalar potential} we use in this work, as in
\cite{RamppJanka2002}-\cite{Dimmelmeier2006} we ``mix together''
equations (\ref{Newt})-(\ref{TOV}), i.e. we describe the equilibrium
configurations using the following equations
\begin{eqnarray}
\label{TOVNewt}
\frac{dP}{dr} &=& -\frac{(\eps+P)(m_{TOV}+4\pi r^3P)}{r(r-2m_{TOV})}\\
\nonumber
\frac{dm_{TOV}}{dr} &=& 4\pi\eps r^2 \\
\nonumber
\frac{d\Phi_{TOV}}{dr} &=& -\frac{1}{\rho}\frac{dP}{dr}
\end{eqnarray}
where the first two are the first two TOV equations (\ref{TOV}), and
the last is the second Newtonian equation (\ref{Newt}).  The potential
$\Phi_{TOV}(r)$ obtained by integrating equations (\ref{TOVNewt}) is
then used in the self-gravity tensor
\begin{eqnarray}
\hat{V}_{ij} = -\int dM \partial_i(\Phi)r_j
\end{eqnarray}
which is needed to find $\hat{V}= \textrm{Tr}(\hat{V}_{ij})$ used in
the self-gravity potential  (\ref{def:V}).

\section{Tidal Disruption Data from Different Models}\label{app:table}
In Table \ref{cfr_data}, we provide the data plotted in Figure
\ref{TBFS}. A polytropic EOS with $n=1$ and $\kappa=1$ is used to
model the star.  Each data set corresponds to one of the panels in
Figure \ref{TBFS}, which is identified by the row indicating the NS
compactness $C$ and its gravitational mass normalised with respect to
the polytropic length scale $R_{poly}=\kappa^{n/2}$. The first column
gives the mass ratio $q$. The remaining three columns are the orbital
angular frequencies at tidal disruption, normalised with respect to
$R_{poly}/10$, resulting from calculations performed with the three
binary models considered: ``Full GR'' indicates data extracted from
Figure 11 of \cite{TBFS08}, ``Pseudo GR'' data obtained with our model
and ``Newtonian'' data obtained with the model of \cite{WL2000} in
which the NS self-gravity is Newtonian. We remind the reader that the
BH is always non rotating (the BHs in \cite{TBFS08} actually have a
small residual angular momentum, see the reference for further
details).

\begin{table}[!h]
\centering
\begin{tabular}{|c||c|c|c|}
\hline
\multicolumn{2}{|c}{$C= 0.1088$} &
\multicolumn{2}{c|}{$M_{NS}/R_{poly}= 0.1136$}\\
\hline
$q$ & Full GR & Pseudo GR & Newtonian \\
\hline
$9$ & $0.88$ & $0.88$ & $0.70$ \\
$8$ & $0.89$ & $0.91$ & $0.71$ \\
$7.5$ & $0.90$ & $0.92$ & $0.72$ \\
$7$ & $0.91$ & $0.94$ & $0.74$ \\
$6$ & $0.93$ & $0.96$ & $0.75$ \\
\hline
\multicolumn{2}{|c}{$C= 0.1201$} &
\multicolumn{2}{c|}{$M_{NS}/R_{poly}= 0.1223$}\\
\hline
$7$ & $0.97$ & $0.99$ & $0.75$ \\
$6.5$ & $0.99$ & $1.00$ & $0.76$ \\
$6$ & $1.00$ & $1.02$ & $0.77$ \\
$5$ & $1.04$ & $1.05$ & $0.80$ \\
$3$ & $1.08$ & $1.16$ & $0.87$ \\
\hline
\multicolumn{2}{|c}{$C= 0.1321$} &
\multicolumn{2}{c|}{$M_{NS}/R_{poly}= 0.1310$}\\
\hline
$6$ & $1.06$ & $1.07$ & $0.79$ \\
$5$ & $1.08$ & $1.11$ & $0.82$ \\
$4$ & $1.11$ & $1.17$ & $0.85$ \\
$3$ & $1.16$ & $1.24$ & $0.89$ \\
\hline
\multicolumn{2}{|c}{$C= 0.1452$} &
\multicolumn{2}{c|}{$M_{NS}/R_{poly}= 0.1395$}\\
\hline
$5$ & $1.21$ & $1.18$ & $0.84$ \\
$4$ & $1.22$ & $1.24$ & $0.87$ \\
$3$ & $1.28$ & $1.33$ & $0.92$ \\
$2$ & $1.31$ & $1.45$ & $0.99$ \\
$1$ & $1.37$ & $1.72$ & $1.17$ \\
\hline
\multicolumn{2}{|c}{$C= 0.1600$} &
\multicolumn{2}{c|}{$M_{NS}/R_{poly}= 0.1478$}\\
\hline
$4.5$ & $1.28$ & $1.29$ & $0.87$ \\
$4$ & $1.31$ & $1.33$ & $0.89$ \\
$3$ & $1.35$ & $1.43$ & $0.94$ \\
\hline
\multicolumn{2}{|c}{$C= 0.1780$} &
\multicolumn{2}{c|}{$M_{NS}/R_{poly}= 0.1560$}\\
\hline
$3.5$ & $1.49$ & $1.50$ & $0.93$ \\
$3$ & $1.53$ & $1.56$ & $0.96$ \\
$2$ & $1.61$ & $1.73$ & $1.04$ \\
\hline
\end{tabular}
\caption{Data plotted in Figure \ref{TBFS}: each subtable corresponds
  to a panel of the figure, which is identified by the stellar
  compactness $C$ and the gravitational mass normalised with respect
  to $R_{poly}= \kappa^{n/2}$. The mass ratio $q$ is given in the
  first column, while the orbital angular frequency calculated at the
  tidal disruption limit resulting from each model considered here ---
  Full-GR, Pseudo GR, Newtonian --- is displayed in the form
  $\Omega_{tide}R_{poly}/10$ in the other three columns.  See the text
  of the appendix for more details.}
\label{cfr_data}
\end{table}


\end{document}